\documentclass[9pt,twocolumn,twoside]{opticajnl}
\journal{opticajournal} % use for journal or Optica Open submissions

\setboolean{shortarticle}{false}
\usepackage{siunitx}

\newcommand{\TE}{\mathrm{TE}}
\newcommand{\TM}{\mathrm{TM}}
\newcommand{\PW}{\mathrm{PW}}
\newcommand{\EW}{\mathrm{EW}}
\newcommand{\kz}{k_\perp}
\newcommand{\kzm}{k^{(m)}_\perp}
\newcommand{\kB}{k_\mathrm{B}}
\renewcommand{\Re}{\operatorname{Re}}
\renewcommand{\Im}{\operatorname{Im}}

\title{Casimir repulsion with biased semiconductors}

\author[1]{Benjamin Spreng}
\author[1]{Calum Shelden}
\author[1]{Tao Gong}
\author[1,*]{Jeremy N. Munday}

\affil[1]{Department of Electrical and Computer Engineering, University of California, Davis, CA 95616, USA}

\affil[*]{jnmunday@ucdavis.edu}

\begin{abstract}
Quantum and thermal fluctuations are fundamental to a plethora of phenomena within quantum optics, including the Casimir effect that acts between closely separated surfaces typically found in MEMS and NEMS devices. Particularly promising for engineering and harnessing these forces are systems out of thermal equilibrium.
Recently, semiconductors with external bias have been proposed to study the nonequilibrium Casimir force.
Here, we explore systems involving moderately biased semiconductors that exhibit strong repulsive Casimir forces, and we determine the effects of bias voltage, semiconductor bandgap energy, and separation for experimentally accessible configurations.
Modes emitted from the semiconductors exert a repulsive force on a near surface that overcomes the attractive equilibrium Casimir force contribution at submicron distances.
For the geometry of two parallel planes, those modes undergo Fabry-P\'erot interference resulting in an oscillatory force behavior as a function of separation.
Utilizing the proximity-force approximation, we predict that the repulsive force exerted on a gold sphere is well within the accuracy of typical Casimir force experiments.
Our work opens up new possibilities of controlling forces at the nano- and micrometer scale with applications in sensing and actuation in nanotechnology.
\end{abstract}

\setboolean{displaycopyright}{false} % Do not include copyright or licensing information in submission.

\begin{document}

\maketitle

\section{Introduction}

Quantum fluctuations are a cornerstone of quantum optics, underlying fundamental nonclassical phenomena such as spontaneous emission and the Lamb shift \cite{milonni_quantum_1994}. 
These zero-point fluctuations also set constraints on the design of nano-scale devices.
For instance, they introduce a fundamental noise floor that limits the sensitivity of quantum optomechanical devices \cite{aspelmeyer_cavity_2014, barzanjeh_optomechanics_2022}.
Moreover, quantum fluctuations together with thermal fluctuations give rise to the Casimir force that typically attract closely spaced macroscopic surfaces at the nanometer to micron scale \cite{Casimir_1948, bordag_advances_2009}. This force leads to the well-known stiction problem in micro- and nanoelectromechanical systems (MEMS and NEMS) that limits their functionality and reliability \cite{chan_quantum_2001, palasantzas_applications_2020}. Systems that exhibit a repulsive Casimir force thus have immense practical implications.

Repulsive Casimir forces have been experimentally demonstrated for interacting surfaces immersed in a fluid \cite{munday_measured_2009}; yet, the quest to replicate this repulsion across air or vacuum, which holds greater relevance for practical applications, presents a more formidable challenge \cite{gong_recent_2020}. At thermal equilibrium, repulsive forces have been predicted for two parallel plates using various materials and conditions. These include magnetic materials \cite{boyer_van_1974, henkel_casimir_2005, geyer_thermal_2010, inui_quantum_2012, shelden_enhanced_2023}, topological insulators \cite{grushin_tunable_2011, martinez_tuning_2013, nie_casimir_2013}, Chern insulators \cite{rodriguez-lopez_repulsive_2014}, and Weyl semimetals \cite{wilson_repulsive_2015}. Additionally, a purely geometry-based repulsive force has been theorized for specific configurations, such as a needle in front of a metallic plate with a hole \cite{levin_casimir_2010}.

Systems out of thermal equilibrium offer another promising avenue for achieving repulsive Casimir forces \cite{antezza_casimirlifshitz_2006, antezza_casimirlifshitz_2008, kruger_nonequilibrium_2011, messina_casimirlifshitz_2011, messina_scatteringmatrix_2011, bimonte_dilution_2011, kruger_trace_2012, noto_casimirlifshitz_2014}. These nonequilibrium systems not only demonstrate potential for repulsion but also exhibit unique characteristics distinct from their equilibrium counterparts. One such feature is an extended force range due to the radiation pressure exerted by photons \cite{antezza_casimirlifshitz_2008}. Furthermore, the force in these systems can vary in sign with changes in separation distance, enabling the creation of stable equilibrium points \cite{kruger_nonequilibrium_2011, bimonte_dilution_2011}. Notably, the forces acting on the involved objects are not always equal in magnitude, and in certain configurations, can even induce self-propelling behaviors \cite{kruger_nonequilibrium_2011}.
A key difference in nonequilibrium systems is the role of material resonances, which become significantly more relevant compared to the broadband dielectric response observed in equilibrium Casimir forces \cite{henkel_radiation_2002, cohen_resonant_2003, bimonte_dilution_2011, kruger_trace_2012}. This opens the possibility of precisely engineering the Casimir force by tuning these material resonances, in conjunction with carefully controlling the temperatures of the surfaces and their surrounding environment.

Experimental evidence for nonequilibrium Casimir forces is notably scarce. To date, there has been only a single experiment involving an ultracold atomic cloud in front of a dielectric substrate \cite{obrecht_measurement_2007}, but no similar studies have been conducted between macroscopic surfaces. A major challenge in these measurements is that nonequilibrium Casimir force contributions are often overshadowed by equilibrium force contributions and only dominate the total force typically at large separations comparable to the thermal wavelength \cite{bimonte_nonequilibrium_2017}, which is approximately $8\,\si{\micro\meter}$ at room temperature. Additionally, observing strong signatures of the nonequilibrium force requires substantial temperature differences, often on the order of hundreds of Kelvin \cite{kruger_trace_2012}. Despite these challenges, for submicron separations and with moderate temperature differences, a differential force measurement setup has been proposed to observe these forces \cite{bimonte_observing_2015}.

Recently, Chen and Fan proposed a novel approach to exploring nonequilibrium Casimir forces, utilizing the concept of a nonzero chemical potential of photons \cite{chen_nonequilibrium_2016}. This phenomenon emerges when a semiconductor is subject to an external bias voltage, causing the electromagnetic fluctuations within the semiconductor to gain a nonzero chemical potential, which is proportional to the applied voltage \cite{landsberg_photons_1981, pwurfel_chemical_1982, feuerbacher_verification_1990}. In their study, Chen and Fan considered a semiconductor sphere with radius of $1\,\si{\micro\meter}$ in front of a semiconductor plane subject to different voltages and found the nonequilibrium Casimir force can exceed its equilibrium counterpart even at nanoscale separations.
Moreover, for relatively moderate biases they predict repulsive forces of up to $0.1\,\si{\pico\newton}$ on the small sphere.

In this paper, we propose and analyze two experimental configurations that lead to large nonequilibrium repulsive Casimir forces using biased semiconductors. Our analysis begins with a system of parallel planes, providing an intuitive physical understanding of the nonequilibrium contributions to the force. We find that a biased semiconductor plate can exert a repulsive force of several millipascals on a gold (Au) plate at submicron separations, which may be of particular interest in view of the anticipated high-precision experiment between parallel planes CANNEX \cite{sedmik_next_2021}. Next, we apply our methods to a sphere positioned in front of the plate using the proximity-force approximation (PFA). Our results reveal that the force on the sphere, particularly for typical sphere radii ranging from $50$ to $150\,\mu$m, is orders of magnitude higher than the forces predicted in the work of Chen and Fan. This difference can be attributed to the increased size and metallic character of the sphere and shows that repulsive nonequilibrium Casimir forces should be observable with current experimental techniques.

This paper is organized as follows: Section~\ref{sec:2} develops the theoretical framework for calculating the nonequilibrium Casimir pressure exerted on one plate facing a second plate consisting of a biased semiconductor. We apply this theory to scenarios reflective of realistic experimental conditions and identify materials with promising characteristics. Additionally, the asymptotic behavior of the nonequilibrium pressure at large separations is explored. Section~\ref{sec:3} extends our theoretical approach to a spherical object near the semiconductor plate, employing the proximity-force approximation. We introduce a formula that is straightforward to compute numerically, and present results using parameters typical of experimental setups in atomic force microscope and MEMS systems. The paper concludes in Sec.~\ref{sec:4}, where we summarize our findings and provide perspectives for future research.

\section{Nonequilibrium Casimir pressure on a plate}
\label{sec:2}
\subsection{Theoretical framework}
\label{sec:2A}
\begin{figure}[ht]
\centering
\mbox{\includegraphics[width=\linewidth]{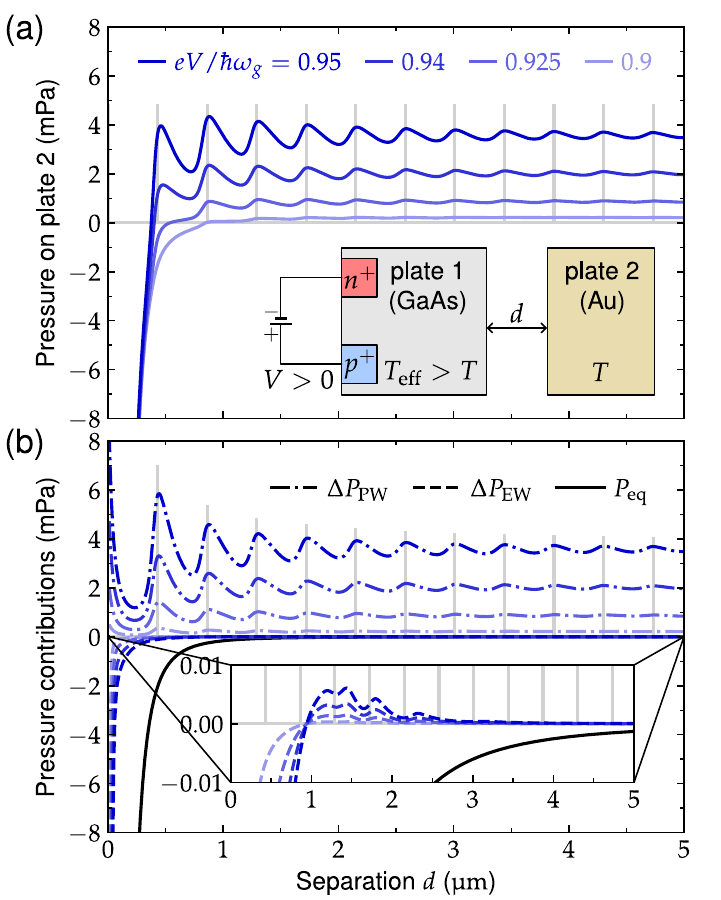}}
\caption{Casimir pressure as a function of separation $d$ exerted on an Au plate in front of a GaAs plate with applied forward-bias $V>0$ at temperature $T=300\,$K. (a) Pressure for relative bias values of $eV/\hbar\omega_g=0.95$, $0.94$, $0.925$ and $0.9$ with bandgap frequency $\omega_g=1.43\,$eV. Inset: Geometry of a semiconductor plate with forward bias in front of a gold plate. (b) Decomposition of the pressure contributions according to \eqref{eq:P_neq} into the equilibrium Casimir pressure $P_\text{eq}$ (solid), propagating wave contribution $\Delta P_\text{PW}$ (dash-dotted), and evanescent wave contribution $\Delta P_\text{EW}$ (dashed). The vertical grey bars indicate integer multiples of the separation at which the PW constructively interfere, i.e. $d=\pi c/\omega_g\approx 430\,$nm. The inset shows a zoom into a smaller section of the vertical axis.}
\label{fig:fig1}
\end{figure}

Consider a configuration where a semiconductor plate with a bias voltage $V$ is positioned at a separation $d$ from another plate, as illustrated in the inset of Fig.~\ref{fig:fig1}(a). Both plates are situated in a vacuum environment with the whole system being held at a constant temperature $T$. We assume that the plates are sufficiently thick to absorb all non-reflected waves. Additionally, their lateral dimensions are considerably larger than the separation between them, allowing us to model the reflection coefficients on their surfaces as those of semi-infinite plates. For the two polarizations $p=\TE$ and $\TM$, these coefficients are given by
\begin{equation}\label{eq:reflection_coefficients}
\begin{aligned}
r^{(m)}_\TE &= \frac{\kz - \kzm}{\kz + \kzm}\,,\\
r^{(m)}_\TM &= \frac{\varepsilon^{(m)} \kz - \kzm}{\varepsilon^{(m)} \kz + \kzm}\,,
\end{aligned}
\end{equation}
where $\kz=\sqrt{\omega^2/c^2 - k^2}$ and $\kzm=\sqrt{\varepsilon^{(m)}\omega^2/c^2 - k^2}$ are the wave vector components perpendicular to the plates in vacuum and inside the plate $m$ with dielectric function $\varepsilon^{(m)}$, respectively. Moreover, $\omega$ is the angular frequency and $k$ the in-plane component of the wave vector.

An applied bias on the semiconductor plate changes the thermal distribution of photons emitted from the plate. While the thermal photon population per mode $n$ for frequencies below the bandgap frequency $\omega_g$ remains unchanged compared to the system without bias, the thermal photon population above the bandgap frequency can be described by the Bose-Einstein distribution \cite{landsberg_photons_1981, pwurfel_chemical_1982}
\begin{equation}\label{eq:n_bias}
n(\omega, T, V) = \frac{1}{\exp((\hbar\omega-eV)/k_B T) - 1}\quad\text{for $\omega > \omega_g$}\,,
\end{equation}
where the term $eV$ is the chemical potential for photons. For a forward-biased plate with $V>0$, \eqref{eq:n_bias} shows that the emission from the plate is increased through the application of the bias and vice versa for a negatively-biased plate. While the following formalism applies also to a negative bias, we will focus on a forward bias here because the increase of emission from the semiconductor plate favors a repulsive force on nearby surfaces through an increase in radiation pressure.
In practice, \eqref{eq:n_bias} is only valid in the spontaneous emission regime, $\hbar\omega_g - eV \ll \kB T$,  in which the chemical potential of photons is much smaller than the bandgap energy with respect to the thermal energy $\kB T$. 

The impact of applying a bias to a semiconductor surface is often interpreted as an effective change in temperature \cite{berdahl_radiant_1985, zhu_nearfield_2019}. An effective temperature $T_\mathrm{eff}$ can be defined by setting $n(\omega_g, T, V) =  n(\omega_g, T_\mathrm{eff}, 0)$.
This substitution leads to the formula $T_\mathrm{eff} = T/[1 - eV/\hbar\omega_g]$, 
showing that a forward bias increases the effective temperature of the semiconductor surface. The concept of effective temperature provides an intuitive way to understand how a bias affects the Casimir interaction in semiconductors. However, it's important to note that this description has its limitations. Specifically, applying a bias alters the photon emission primarily around the bandgap frequency. This emission is different from the broader spectral changes that would occur with an actual increase in temperature. Therefore, while the effective temperature description is helpful for gaining an initial understanding, it does not fully capture the nuances of how bias impacts the Casimir interaction.

We now focus on the Casimir pressure exerted on the plate in front of the biased semiconductor. It is convenient to split the net Casimir pressure on that plate into two parts. The first part is the contribution stemming from the modes present at thermal equilibrium without bias, and it is given by the Lifshitz formula \cite{Lifshitz_SovPhysJETP_1956, Dzyaloshinskii_AdvPhys_1961}
\begin{multline}\label{eq:lifshitz}
P_\mathrm{eq}(d, T) = \\
 - \frac{\kB T}{\pi}\sum_{n=0}^\infty{}' \int_0^\infty \mathrm{d}k\, k\kz
\sum_{p=\TE,\TM} \frac{r_p^{(1)}r_p^{(2)}\exp(-2\kz d)}{1 - r_p^{(1)}r_p^{(2)}\exp(-2\kz d)}\,,
\end{multline}
where the reflection coefficients \eqref{eq:reflection_coefficients} and $\kz$ are evaluated at the imaginary Matsubara frequencies $\omega=i\xi_n$ with $\xi_n=2\pi n \kB T/\hbar$ and the primed sum indicates that the $n=0$ term is weighted with a factor of $1/2$. We follow the convention that a negative sign of the pressure corresponds to attraction and a positive sign to repulsion.

The second contribution to the net Casimir pressure on plate 2 stems from the excess of modes emitted from the semiconductor plate due to the presence of the forward bias. This excess of modes is only present for frequencies above the bangap frequency, and the thermal photon population of the excess of modes can be described by the difference
\begin{equation}
\Delta n(\omega, T, V) = n(\omega, T, V) - n(\omega, T, V=0)\,.
\end{equation}

For the nonequilibrium Casimir pressure contribution due to the bias, we further distinguish between contributions from propagating waves (PW) and evanescent waves (EW) and denote the corresponding Casimir pressures as $\Delta P_\PW(d, T, V)$ and $\Delta P_\EW(d, T, V)$, respectively.
To find explicit formulas we can make use of existing formulas that have been worked out in Ref.~\citep{antezza_casimirlifshitz_2008} based on Rytov's theory of fluctuating electromagnetism \citep{rytov_principles_1989}. Equations (62) and (63)  in Ref.~\citep{antezza_casimirlifshitz_2008} describe the Casimir pressure for a system where only one half-space is at a temperature $T$ and the other is at zero temperature. Replacing the factor $[\exp(\hbar\omega/\kB T)-1]^{-1}$ appearing in their formulas by $\Delta n(\omega, T, V)$, we find
\begin{multline}\label{eq:deltaP_PW}
\Delta P_\PW(d, T, V) = \frac{\hbar}{4 \pi^2} \int_{\omega_g}^\infty d\omega\, \Delta n(\omega, T, V)\\\times \int_0^{\omega/c} dk\,k\kz \sum_{p=\TM,\TE}\frac{(1 - \vert r_p^{(1)}\vert^2)(1 + \vert r_p^{(2)}\vert^2)}{\vert D_p \vert^2}\,,
\end{multline}
and
\begin{multline}\label{eq:deltaP_EW}
\Delta P_\EW(d, T, V) = -\frac{\hbar}{\pi^2} \int_{\omega_g}^\infty d\omega\, \Delta n(\omega, T, V)\\ \times \int_{\omega/c}^\infty dk\,k \Im \kz e^{-2 d \Im \kz} \sum_{p=\TM,\TE}\frac{\Im(r_p^{(1)})\Re(r_p^{(2)})}{\vert D_p \vert^2}\,,
\end{multline}
where $D_p = 1 - r^{(1)}_p r^{(2)}_p \exp(2i \kz d)$.

Equation~(\ref{eq:deltaP_PW}) offers a direct physical interpretation.
The photon flux emitted from plate 1 can be described by Plack's law of radiation where the thermal photon population is replaced by $\Delta n(\omega, T, V)$ and its emissivity is accounted for by a factor $(1-\vert r_p^{(1)}\vert^2)/2$ for each polarization.
Within the Fabry-P\'erot cavity formed by the two plates, the fields are formed by interference of emitted modes with those reflected multiple times between the plates. Consequently, the electric fields between the plates are modified by a factor of $1/D_p$, leading to a modification of the photon flux by $1/\vert D_p\vert^2$. 
Incident photons on plate 2 are either absorbed or reflected transferring a total momentum of $(1-\vert r^{(2)}_p\vert^2)\hbar\kz$ on plate 2. Finally by integrating over all modes, we obtain expression \eqref{eq:deltaP_PW}. Because the momentum transferred on plate 2 is always positive, the nonequilibrium contributions due to propagating waves are inherently repulsive across all separations. However, a physical interpretation of \eqref{eq:deltaP_EW} is more complex, as it involves evanescent waves, and its contribution can be either attractive or repulsive in general.

Putting the contributions from the equilibrium Casimir force \eqref{eq:lifshitz} and the nonequilibrium contributions due to the bias \eqref{eq:deltaP_PW} and \eqref{eq:deltaP_EW} all together, the net Casimir pressure on plate 2 can then be expressed as
\begin{equation}\label{eq:P_neq}
P^{(2)}_\text{neq}(d, T, V) = P_\mathrm{eq}(d, T) + \Delta P_\PW(d, T, V) + \Delta P_\EW(d, T, V)\,.
\end{equation}
Notice that the total Casimir pressure on plate 1, $P^{(1)}_\text{neq}(d, T, V)$, is different from \eqref{eq:P_neq} because the pressure due to emission of modes into the space facing away from plate 2 also needs to be taken into account.

\subsection{Application}

As a first example, we illustrate the properties of the net Casimir pressure \eqref{eq:P_neq} and its constituent contributions by considering a GaAs plate with bandgap of $1.43\,$eV in front of a gold plate. We model the dielectric functions of all materials appearing in this manuscript according to tabulated data from Ref.~\cite{palik_handbook_1985} and extrapolate the dielectric function of gold towards zero frequency using a Drude model with a plasma frequency of $9\,$eV and damping frequency of $35\,$meV. Moreover, the temperature is assumed to be  $T=300\,$K throughout the paper.

In Fig.~\ref{fig:fig1}(a), we illustrate how the Casimir pressure on plate 2, as described by \eqref{eq:P_neq}, varies with separation for a constant applied bias across the GaAs plate. To allow for a comparison of our results for different semiconductor materials, we express the bias relative to the bandgap frequency in terms of the dimensionless quantity $eV/\hbar\omega_g$. The pressure shifts from attractive at shorter separations to repulsive at larger ones, exhibiting slowly decaying oscillations around a stable value. Notably, the transition point from attractive to repulsive pressure decreases as the bias increases, occurring at separations of $380\,$nm, $400\,$nm, $540\,$nm, and $850\,$nm for the largest to smallest bias values, respectively. Furthermore, both the magnitude of the asymptotic constant repulsive pressure and the oscillation amplitude are observed to increase with an increase in the bias.

In Fig.~\ref{fig:fig1}(b), we present the individual components contributing to the net Casimir pressure, as detailed on the right-hand side of \eqref{eq:P_neq}. This figure highlights that the repulsive nature and oscillatory behavior of the pressure are primarily due to the PW contribution, as given in \eqref{eq:deltaP_PW}. These effects are largely associated with modes in a narrow frequency band around the bandgap. The oscillations can be attributed to Fabry-P\'erot interferences, particularly evident as the pressure peaks closely align with resonant separation distances, marked by vertical grey bars.

While the PW contribution remains repulsive and bounded, consistent with the concept of radiation pressure, the EW contribution exhibits a shift from attraction to repulsion at approximately $0.94\,\si{\micro\meter}$ and diverges at short separations, as shown in the inset. Conversely, the equilibrium pressure component, $P_\text{eq}$, is consistently attractive across all separations and becomes the dominant force at shorter distances, where it also diverges. The shift in the net Casimir pressure from attraction to repulsion can be understood as a result of the interplay between the diverging attractive Casimir pressure and the repulsive nonequilibrium PW component.

The increasing magnitude of the nonequilibrium pressure contributions \eqref{eq:deltaP_PW} and \eqref{eq:deltaP_EW} with increasing bias can be attributed to the increase of emitted photons from the semiconductor plate. 
However, note that these pressure contributions exhibit a logarithmic divergence, following $-\ln((\hbar\omega_g-eV)/ \kB T)$ as the relative bias approaches unity. 
This divergence marks the semiconductor's transition into a lasing mode, suggesting that the applied formulas remain valid as long as the condition $\hbar\omega_g-eV \ll \kB T$ is met. In the given example with GaAs, this condition is indeed fulfilled because $\hbar\omega_g-eV = 71\,$meV for the largest bias, while $\kB T=26\,$meV at room temperature.

\begin{figure}[ht]
\centering
\mbox{\includegraphics[width=\linewidth]{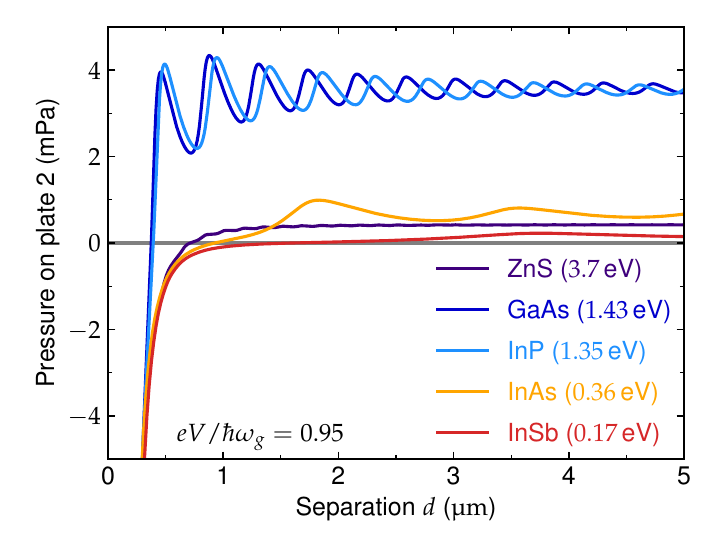}}
\caption{Casimir pressure as a function of separation $d$ exerted on a Au plate in front of a semiconductor plate of a given material (see legend) at a fixed relative bias $eV/\hbar\omega_g=0.95$. The values in the legend indicate the bandgap frequency of each semiconductor, respectively.}
\label{fig:fig2}
\end{figure}

In Fig.~\ref{fig:fig2}, we compare the Casimir pressure on the gold plate for various choices of semiconductor materials for the biased plate. Note that for ZnS, we assume a cubic lattice structure with isotropic optical data from \citep{palik_handbook_1985}. The relative bias is fixed to $eV/\hbar\omega_g=0.95$ for all curves. Overall, the curves share similar features of an attractive force at short separations turning to a repulsive pressure that oscillates around a constant value for larger separations. The periodicity of the oscillations is consistent with the Fabry-P\'erot resonance condition discussed above. Interestingly, the large-distance repulsive pressure seems to indicate a nonmonotonic relationship with the bandgap frequency. It is largest for GaAs with a pressure of $3.6\,$mPa followed by InP with $3.5\,$mPa. The other materials have a much smaller asymptotic value of $0.7$, $0.4$ and $0.2\,$mPa for InAs, ZnS and InSb, respectively. Our results for the repulsive Casimir pressure values fall well within the anticipated sensitivity of the CANNEX experiment of $10^{-9}\,$mPa \cite{sedmik_next_2021}.

\subsection{Pressure at large separations}

In the previous examples, we have seen that for a forward-biased semiconductor plate the Casimir pressure becomes repulsive at larger separations. In this section, we analyze the behavior of the net Casimir pressure exerted on plate 2 and its constituent contributions at large separations. 
Here, 'large separation' is defined not solely as a measure exceeding the thermal wavelength $\lambda_T = \hbar c/\kB T$, which equals $7.6\,\si{\micro\meter}$ at room temperature, but also in relation to the wavelength associated with the bandgap frequency.

The behavior of the equilibrium Casimir pressure at large separations $d\gg \lambda_T$ is well known \citep{bordag_advances_2009}. Only the zero-frequency Matsubara contribution in \eqref{eq:lifshitz} plays a role in that case. At vanishing frequencies, the Fresnel coefficients in \eqref{eq:reflection_coefficients} become $r^{(m)}_\TM=(\varepsilon^{(m)}(0)-1)/(\varepsilon^{(m)}(0)+1)$ and $r^{(m)}_\TE=0$ for dielectrics and $r^{(m)}_\TM=1$ and $r^{(m)}_\TE=0$ for metals described by a Drude model. Contributions from TE polarization thus do not contribute at large separations for both cases. Performing the integral over $k$ in \eqref{eq:lifshitz} then yields \cite{bordag_advances_2009}
\begin{equation}
P_\text{eq}(d,T) \approx -\frac{\kB T}{8\pi d^3}\operatorname{Li}_3(r_\TM^{(1)} r_\TM^{(2)})\,,
\end{equation}
where the zero-frequency limit is taken for the reflection coefficients and $\operatorname{Li}_s$ is the polylogarithm of order $s$ \cite{NIST:DLMF}.

To find the asymptotic behavior of the EW nonequilibrium contribution \eqref{eq:deltaP_EW}, we make use of the multi-reflection formula \citep{antezza_casimirlifshitz_2008}
\begin{equation}\label{eq:EW_expansion}
\frac{\exp(-2q d)}{\vert 1- \mathcal{R} \exp(-2 q d)\vert^2} = \sum_{s=1}^\infty \frac{\Im \mathcal{R}^s}{\Im \mathcal{R}} \exp(-2sq d) 
\end{equation}
with $\mathcal{R}= r^{(1)}_p r^{(2)}_p$ and $q = \Im\kz$.
It is then convenient to change the integration variable from $k$ to $q$. For large separations, $d\gg 1/q$, the contributions of the integrand around $q=0$ are most important. Expanding the reflection coefficients for small $q \ll \omega/c$, we find
\begin{equation}
r_p^{(m)} = -1 + \frac{2 i q c \gamma_p^{(m)}}{\omega} + O(q^2)
\end{equation} 
with $\gamma_\TE^{(m)} = 1/\sqrt{\varepsilon^{(m)} - 1}$ and $\gamma_\TM^{(m)} = \varepsilon^{(m)}/\sqrt{\varepsilon^{(m)} - 1}$. Further using that $\Im \mathcal{R}^s/\Im \mathcal{R} = s$ to leading order in $q$ and performing the integral over $q$ explicitly, we find our final result
\begin{equation}\label{eq:P_EW_asymptotics}
\Delta P_\EW(d, T, V) \approx \frac{3 \zeta(3)\hbar c }{4 d^4 \pi^2} \int_{\omega_g}^\infty \frac{d\omega}{\omega}\,\Delta n(\omega, T, V) \sum_{p=\TE,\TM} \Re(\gamma_p^{(1)})
\end{equation}
valid for $d \gg c/\omega_g$. Note that the large-separation result of \eqref{eq:P_EW_asymptotics} for $\Delta P_\EW$ is primarily a function of the dielectric properties of plate 1, as characterized by $\gamma_p^{(1)}$, while the dielectric function of plate 2 only contributes to subleading terms not considered here.
For large distances, the contribution $\Delta P_\EW(d, T, V)$ thus decays by a power in the separation faster than the equilibrium contribution $P_\text{eq}$, which is consistent with the numerical results depicted in Fig.~\ref{fig:fig1}(b).

We next examine the asymptotic behavior of the PW nonequilibrium contribution \eqref{eq:deltaP_PW} for large separations. To this end, it is convenient to employ the expansion \citep{antezza_casimirlifshitz_2008}
\begin{multline}\label{eq:PW-expansion}
\frac{1}{\vert 1- \mathcal{R} \exp(2i \kz d)\vert^2} = \\
\frac{1}{1- \vert \mathcal{R} \vert^2} \left( 1 + 2 \Re\sum_{s=1}^\infty \mathcal{R}^s \exp(2is \kz d) \right)\,,
\end{multline}
where again $\mathcal{R}= r^{(1)}_p r^{(2)}_p$.
Note that in Ref.~\citep{antezza_casimirlifshitz_2008} this expansion was interpreted as a multi-reflection expansion for which the first term corresponds to radiation that passes the cavity only once without being reflected.
However, our discussion in Sec.~\ref{sec:2A} suggests a different interpretation.
As we have discussed above, the factor $1/\vert D_p\vert^2$ being expanded here describes the interference of the emitted mode with those reflected an arbitrary number of times between the plates. Thus, for a mode that passes the cavity only once without being reflected, this factor should be equal to one and not depend on the reflection coefficients.
A more suitable physical interpretation is that \eqref{eq:PW-expansion} is an expansion in terms of phases accumulated by the interfering modes between the two plates.
The first term in the parenthesis then corresponds to a ray-optical limit for which interference is neglected and no phase factor is accumulated, and the second term sums over all possible non-vanishing phases.

Applying the expansion of \eqref{eq:PW-expansion} to \eqref{eq:deltaP_PW}, we can split the nonequilibrium PW contribution into a separation independent and a separation dependent contribution and write
\begin{equation}\label{eq:PW-split}
\Delta P_\PW(d, T, V) = \Delta P_{\PW,0}(T, V) +  \Delta P_{\PW,1}(d, T, V)\,,
\end{equation}
where the first term corresponds to the separation independent ray-optical limit and reads
%\begin{multline}\label{eq:DeltaP_PW,0}
%\Delta P_{\PW,0}(T, V) = \frac{\hbar}{4 \pi^2} \int_{\omega_g}^\infty d\omega\, \Delta n(\omega, T, V)\\\times \int_0^{\omega/c} dk\,k\kz \sum_{p=\TM,\TE}\frac{(1 - \vert r_p^{(1)}\vert^2)(1 + \vert r_p^{(2)}\vert^2)}{1 - \vert r_p^{(1)} r_p^{(2)} \vert^2}\,.
%\end{multline}
%alternative version
\begin{multline}\label{eq:DeltaP_PW,0}
\Delta P_{\PW,0}(T, V) = \frac{\hbar}{4 \pi^2} \int_{\omega_g}^\infty d\omega\, \Delta n(\omega, T, V)\\\times\int_0^{\omega/c} dk\,k\kz [f_\TE(\omega, k) + f_\TM(\omega, k)]
\end{multline}
with
\begin{equation}
f_p(\omega, k) = \frac{(1 - \vert r_p^{(1)}\vert^2)(1 + \vert r_p^{(2)}\vert^2)}{1 - \vert r_p^{(1)} r_p^{(2)} \vert^2}\,.
\end{equation}
It is this term that describes the large-separation limit around which the net Casimir pressure on the gold plate oscillates in Figs.~\ref{fig:fig1} and \ref{fig:fig2}.

The second term on the right-hand side of \eqref{eq:PW-split} captures the dependence on separation through the sum over nonvanishing phase factors in the expansion \eqref{eq:PW-expansion}. It is again convenient to change the integration variable; this time from $k$ to $k_\perp$. For separations much larger than the bandgap wavelength, $d \gg \pi c/\omega_g$, an asymptotic expansion of the integral over $\kz$ can be obtained by means of a partial integration. The most important contribution to the integral then comes from the vicinity of $\kz = 0$ corresponding physically to modes normal to the plates' surfaces. We thus find
\begin{multline}\label{eq:dP_PW,1_intermediate}
\Delta P_{\PW,1}(d, T, V) \approx -\frac{\hbar}{2 c^2 \pi^2 d} \int_{\omega_g}^\infty d\omega\, \omega^2 \Delta n(\omega, T, V) \\
\times f(\omega) \Im \ln(1 - r^{(1)} r^{(2)} e^{2i\omega d/c})
\end{multline}
with $f(\omega) \equiv f_\TE(\omega,0)=f_\TM(\omega,0)$  and the reflection coefficients at normal incidence $r^{(m)}= (1 - \sqrt{\varepsilon^{(m)}})/(1 + \sqrt{\varepsilon^{(m)}})$.

For separations large compared to the thermal wavelength $d\gg \lambda_T$, we can further approximate the integral over frequencies in \eqref{eq:dP_PW,1_intermediate} by employing another partial integration. We then finally obtain
\begin{multline}\label{eq:dP_PW,1_final}
\Delta P_{\PW,1}(d, T, V) \approx \\\frac{\hbar \omega_g^2}{4 c \pi^2 d^2} \Delta n(\omega_g, T, V)  f(\omega_g)\Re \operatorname{Li}_2( r^{(1)} r^{(2)} e^{2i\omega_g d/c})
\end{multline}
valid for $d\gg \lambda_T, \pi c/\omega_g$, where the coefficients $r^{(m)}$ are evaluated at $\omega_g$. Equation~(\ref{eq:dP_PW,1_final}) exhibits the expected oscillatory pressure behavior with respect to separation as indicated by the phase factor appearing in the argument of the polylogarithm. Moreover, our calculations confirm that those oscillations are long range as their decay is proportional to $d^{-2}$ and thus by a power in the separation slower than the equilibrium pressure contribution $P_\text{eq}$.

At large separations, we have thus shown that the net Casimir pressure on the plate 2 is dominated by the PW nonequilibrium contribution $\Delta P_\text{PW}$, which asymptotically approaches the constant value $\Delta P_\text{PW,0}$ given by \eqref{eq:DeltaP_PW,0}. To understand how $\Delta P_\text{PW,0}$ depends on bias and bandgap frequency, we study the ideal case of a perfectly emitting semiconductor plate with $\vert r_p^{(1)} \vert^2 \equiv 0$ paired with a perfectly reflecting plate 2 with $\vert r_p^{(2)} \vert^2 \equiv 1$. In practice, only the dielectric properties of the plates in a narrow band above the bandgap frequency contribute to the nonequilibrium Casimir pressure. An emitting semiconductor plate that is well-absorbing paired with a surface for plate 2 that is well-reflecting around this narrow band can thus be considered as ideal. In fact, the GaAs-Au system depicted in Fig.~\ref{fig:fig1} closely approximates this ideal scenario. Across all biases considered, the values of $\Delta P_\text{PW,0}$ for the GaAs-Au system are consistently at 98\% of those corresponding to the theoretically ideal system.

\begin{figure}[t]
\centering
\mbox{\includegraphics[width=\linewidth]{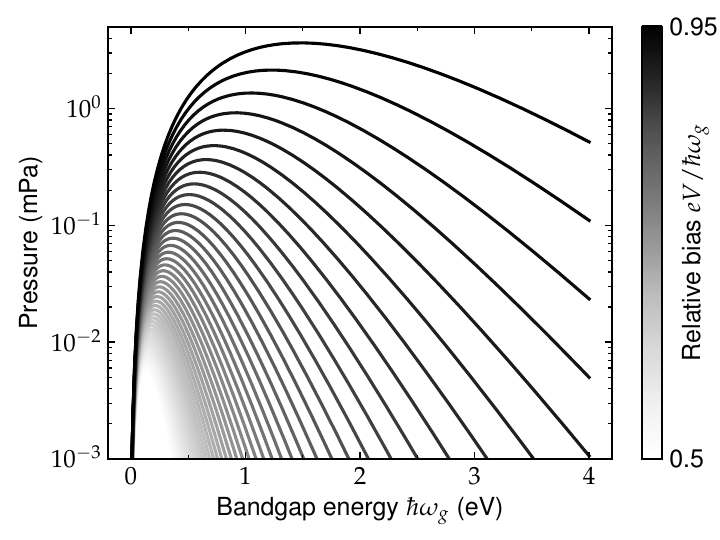}}
\caption{Constant Casimir pressure contribution \eqref{eq:DeltaP_PW,0} for a perfectly emitting plate 1 with $\vert r_p^{(1)} \vert^2 \equiv 0$ and perfectly reflecting plate 2 with $\vert r_p^{(2)} \vert^2 \equiv 1$ as a function of bandgap energy $\hbar\omega_g$ for a fixed relative bias $eV/\hbar\omega_g$. From top to bottom, the curves correspond to a relative bias of $0.95$ decrementing by $0.01$ for each subsequent curve.}
\label{fig:fig3}
\end{figure}

In Fig.~\ref{fig:fig3}, the constant Casimir pressure contribution of \eqref{eq:DeltaP_PW,0}, assuming an ideal system consisting of a perfect emitter and a perfect reflector, is shown as a function of bandgap energy $\hbar\omega_g$ for various fixed values of the relative bias $eV/\hbar\omega_g$. The pressure increases monotonically with the bias as expected. It is, however, nonmonotonic with respect to the bandgap energy resulting in a maximum at an intermediate value of the bandgap energy.
This nonmonotonicity can be understood by first noting that each curve corresponds to the same effective temperature $T_\mathrm{eff}$. The emitted spectrum from the biased semiconductor surface is narrowly peaked near the bandgap energy, and this peak value matches that of a thermal body with temperature $T_\mathrm{eff}$ at $\hbar\omega_g$. As a consequence, the bandgap energy of the pressure maxima corresponds to the energy associated with the maxima in the thermal spectrum. This result further explains the shift of the pressure maxima to higher bandgap frequencies with increasing relative bias.
 
Our results thus offer an explanation of why the net Casimir pressure for the different semiconductor materials oscillates around different values for large separations in Fig.~\ref{fig:fig2}. With the upper-most curve in Fig.~\ref{fig:fig3} corresponding to the same relative bias underlying Fig.~\ref{fig:fig2}, we see that the GaAs and InP systems yield the highest pressure as their bandgap energy aligns closely with the maximum of the curve. The pressure for the other systems is much smaller, because their bandgap energies are much smaller (InAs, InSb) or much larger (ZnS).

\section{Nonequilibrium Casimir force on a sphere}
\label{sec:3}
Most experiments measuring the Casimir force are carried out between a sphere and a plate to avoid the problem of parallelism.
The sphere radius $R$ is typically in the range of $50\,\si{\micro\meter}$ to $150\,\si{\micro\meter}$, while the separation $d$ between them is in the submicron to several microns range \cite{klimchitskaya_recent_2020, gong_recent_2020, bimonte_measurement_2021}.
With the sphere radius being much larger than the separation, numerical exact calculations in the sphere-plane geometry are challenging, and results are so far only available in thermal equilibrium \cite{hartmann_plasma_2017, spreng_planewave_2020}.
However, for large sphere radii, the Casimir force can be well approximated by the Derjaguin approximation also known as the proximity-force approximation (PFA) \cite{derjaguin_theorie_1992, schoger_casimir_2022}. Within the PFA, the Casimir force is obtained by averaging the Casimir pressure of infinitesimal plates across local distances between the surfaces.
Placing the coordinate system on the plate surface, such that the $z$-axis is normal and connects with the sphere center, the local distances in the sphere-plate geometry can described by the function
\begin{equation}\label{eq:local_distance_function}
H(r)=d+ R - \sqrt{R^2 - r^2}\,,
\end{equation}
where $r=\sqrt{x^2+y^2}$ is the distance from the $z$-axis.

Applying our theoretical framework for the nonequilibrium Casimir pressure on a plate developed in Sec.~\ref{sec:2} to the sphere-plate system using the PFA, we can express the Casimir force on the sphere as
\begin{equation}\label{eq:derjaguin}
F^\text{PFA}(d, T, V) = 2\pi \int_0^R r\mathrm{d}r\,P_\text{neq}^{(2)}(H(r), T, V)\,,
\end{equation}
where $P_\text{neq}^{(2)}$ is given in \eqref{eq:P_neq}, and the factor of $2\pi$ is a result of the rotational symmetry of the geometry around the $z$-axis.
Due to the long-range oscillatory behavior of $P_\text{neq}^{(2)}$ a direct numerical evaluation of \eqref{eq:derjaguin} is impracticable as too many evaluations of the function $P_\text{neq}^{(2)}$ are needed. In the following, we thus work out an approximation of formula \eqref{eq:derjaguin} that is numerically easier to compute while still maintaining its accuracy for $d\ll R$.

In view of \eqref{eq:P_neq}, we can apply formula \eqref{eq:derjaguin} term by term and recast the result to the analogous form
\begin{equation}\label{eq:F_PFA}
F^\text{PFA}(d, T, V) = F^\text{PFA}_\text{eq}(d, T) + \Delta F^\text{PFA}_\text{PW}(d, T, V) + \Delta F^\text{PFA}_\text{EW}(d, T, V)\,.
\end{equation}
It is well-known that the equilibrium contribution can be expressed as
\begin{equation}\label{eq:PFAeq}
F^\text{PFA}_\text{eq}(d, T) = 2\pi R \mathcal{F}(d, T)\,,
\end{equation}
where $\mathcal{F}$ is the free energy per unit area for two plates, which is obtained from \eqref{eq:lifshitz} by finding its negative anti-derivative with respect to $d$.

Except for the separation-independent Casimir pressure contribution, which is multiplied by $\pi R^2$ through integration over the interacting surface area in \eqref{eq:derjaguin}, the remaining contributions can be cast into a similar form as \eqref{eq:PFAeq}. Because the Casimir interaction due to those remaining terms falls off with increasing separation, the important part of those contributions comes from the vicinity of $r=0$ around which the local-distance function of \eqref{eq:local_distance_function} can be approximated to $H(r)\approx d + r^2/(2R)$. Employing expansions \eqref{eq:EW_expansion} and \eqref{eq:PW-expansion}, the integrals over $r$ in \eqref{eq:derjaguin} are Gaussian integrals, which can be evaluated straightforwardly.

We then find the explicit formulas
\begin{equation}\label{eq:deltaF_PFA,PW}
\Delta F^\text{PFA}_\PW(d, T, V) = \pi R^2 \Delta P_{\PW, 0}(T, V) + 2\pi R \Phi_\text{PW}(d, T, V)
\end{equation}
and
\begin{equation}
\Delta F^\text{PFA}_\EW(d, T, V) = 2\pi R \Phi_\text{EW}(d, T, V)\,,
\end{equation}
where $\Phi_\text{PW}$ is the negative anti-derivative with respect to $d$ of $\Delta P_\text{PW,1}$ defined through \eqref{eq:PW-split} and reads
\begin{multline}
\Phi_\text{PW}(d, T, V) = \frac{\hbar}{2 \pi} \int_{\omega_g}^\infty d\omega\, \Delta n(\omega, T, V)\\\times \int_0^{\omega/c} dk\,k \sum_{p=\TM,\TE}f_p(\omega, k)\Im\ln(1-r_p^{(1)}r_p^{(2)}\exp(2i\kz d))\,,
\end{multline}
and similarly, $\Phi_\text{EW}$ is the negative anti-derivative of $\Delta P_\text{EW}$, which is given by
\begin{multline}
\Phi_\text{EW}(d, T, V) = -\frac{\hbar}{2\pi^2}\int_{\omega_g}^\infty d\omega\, \Delta n(\omega, T, V)\\
\times \int_{\omega/c}^\infty dk\,k \sum_p \frac{\Im (r_p^{(1)})\Re (r_p^{(2)})}{\Im(r_p^{(1)} r_p^{(2)})}\\
\times\Im \ln(1-r_p^{(1)}r_p^{(2)}\exp(-2\Im\kz d))\,.
\end{multline}

\begin{figure}[t]
\centering
\mbox{\includegraphics[width=\linewidth]{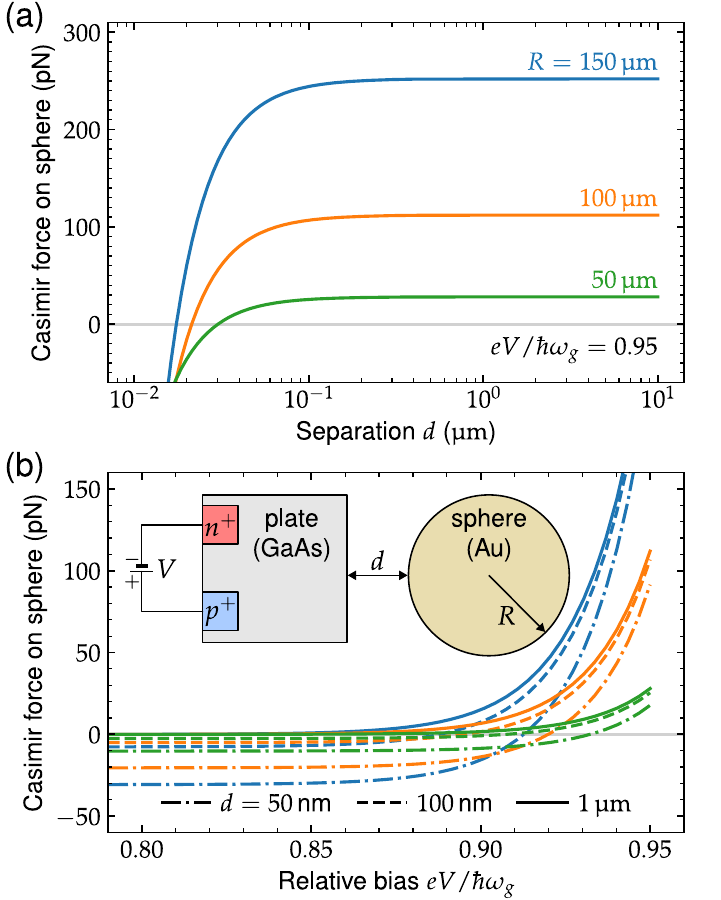}}
\caption{Casimir force within the proximity-force approximation exerted on an Au sphere with radius $R=50\,\si{\micro\meter}$, $100\,\si{\micro\meter}$, $150\,\si{\micro\meter}$ interacting with a GaAs plate as a function of (a) separation at fixed relative bias $eV/\hbar\omega_g=0.95$, and (b) as a function of relative bias at fixed separations of $d=50\,$nm (dash-dotted), $100\,$nm (dashed) and $1\,\si{\micro\meter}$ (solid). Inset: Geometry of a GaAs plate with bias and an Au sphere. }
\label{fig:fig4}
\end{figure}

We thus have found formulas for the Casimir force on a sphere that are of similar numerical complexity to those found for the plate-plate geometry in Sec.~\ref{sec:2}.
In Fig.~\ref{fig:fig4}, we apply the results to the system of an Au sphere in front of a biased GaAs plate, as depicted by the inset.
Panel (a) illustrates the Casimir force exerted on spheres with radii of $R=50\,\si{\micro\meter},$ $100\,\si{\micro\meter}$ and $150\,\si{\micro\meter}$, under a fixed relative bias of $eV/\hbar\omega_g = 0.95$. Similar to the results obtained for planes, the Casimir force transitions from attraction to repulsion at short separations. The point of transition occurs at $17\,$nm, $21\,$nm, and $30\,$nm for the largest to smallest sphere radius, respectively, and is thus at considerably shorter distances than those found in the plane-plane geometry and also those reported by Chen and Fan \cite{chen_nonequilibrium_2016}.
Moreover, the repulsive force observed here ranges between $30\,\si{\pico\newton}$ and $250 \,\si{\pico\newton}$ for the smallest and largest spheres, respectively. The magnitude of these forces is several orders greater than the repulsive force of $0.1\,$pN found by Chen and Fan.

The general behavior of these force curves can be understood as a competition between mainly two contributions in \eqref{eq:F_PFA}. 
At large separations, the constant term in \eqref{eq:deltaF_PFA,PW} dominates, which is proportional to $R^2$, whereas at shorter separations, the diverging equilibrium term \eqref{eq:PFAeq} becomes significant, albeit only multiplied by $R$.
Thus, for increasing sphere radii, the Casimir force increases while the transition separation decreases.

Additionally, the force here does not exhibit notable oscillations around the large-distance limit. This absence of oscillations can be attributed to the fact that the sphere does not form an efficient Fabry-P\'erot cavity with the semiconductor plane and thus interference effects are much weaker. Mathematically, the missing oscillations result from the averaging effect of the integration in \eqref{eq:derjaguin} on the oscillations present in the plate-plate pressure. While weak oscillations in the Casimir force on the sphere still exist, they are too subtle to be discernible at the scale of the figure.

In panel (b), we show the force on the gold sphere as a function of the relative bias while the separation to the GaAs plate is kept fixed at $d=50\,$nm, $100\,$nm and $1\,\si{\micro\meter}$. Regardless of separation and sphere radius, the force increases with increasing bias, as expected. 
Interestingly, our results show that at shorter separations, the force can be tuned from attractive to repulsive by changing the bias alone.
We also note that our results for the magnitude of the Casimir force fall well within the accuracy range of AFM and MEMS measurement techniques.

\section{Conclusion}
\label{sec:4}

In this work, we have presented results detailing the nonequilibrium Casimir force involving semiconductors with externally applied bias. We developed a theoretical framework for the force exerted on a planar surface near a biased semiconductor plate.
The nonequilibrium force contribution overcomes the attractive equilibrium force contribution at submicron separations, and results in a strong repulsive force even for relatively moderate biases.
The repulsive pressure undergoes oscillations as a function of separation due to Fabry-P\'erot interference of modes emitted from the biased semiconductor surface.
Making use of the PFA, we find that the nonequilibrium force on gold spheres, as used in typical experiments, is on the order of tens to hundreds of piconewtons, and thus very well within the sensitivity of state-of-the-art experimental measurement techniques. Further, these results provide a new avenue for tailoring forces on the nano- and microscale, which could have applications in novel optomechanic systems or MEMS devices.

\begin{backmatter}
\bmsection{Funding} The authors acknowledge financial support from the Defense Advanced Research Projects Agency (DARPA) QUEST program Grant No. HR00112090084.

% \bmsection{Acknowledgments}

\bmsection{Disclosures} The authors declare no conflicts of interest.

\bmsection{Data availability} Data underlying the results presented in this paper are available on Zenodo see Ref.~\cite{spreng_zenodo_2024}.

\end{backmatter}

% Bibliography
\bibliography{references}

\end{document}